\documentclass{article}
\usepackage{graphicx} 
\usepackage{amsmath}
\usepackage{authblk}
\usepackage[a4paper, left=1.5in, right=1.5in, top=1.5in, bottom=1.5in]{geometry}

\title{High-Q non-invasive Glucose Sensor using Microstrip Line Main Field and Split Ring Resonator}

\author[1]{Brandon Kaiheng Tay}
\author[1]{Saumitra Kapoor}
\author[2]{Wenwei Yu}
\author[1]{Shao Ying Huang}

\affil[1]{Singapore University of Technology and Design, 487372, Singapore}
\affil[2]{University of Chiba, 1-33 Yayoichō, Inage Ward, Chiba, 263-8522, Japan}

\date{March 2025}

\begin{document}

\maketitle

\begin{abstract}
A high-Q sensor integrating microstrip line (MLIN) main field and split ring resonators is presented for non-invasive glucose sensing. The proposed sensor combines the field-focusing effects of split ring resonators with the enhanced field-substrate interaction properties of the MLIN main field, using the reflection coefficient (S11) of an open-ended MLIN with the finger as the substrate and operating at 750 MHz and 1.5 GHz. The permittivity of blood inside the finger depends on the glucose concentration, which in turn affects the S11 of the system. Sensor geometry was optimized using Method-of-Moments simulation before the sensor was fabricated and validated on standard solutions of glucose concentrations between 0 to 126 mg/dL within the physiological range, and a human test subject. In both experiments, a near inverse-linear relationship between the S11 peak magnitude and the glucose concentration was observed, demonstrating the sensitivity of the proposed sensor for detecting changes in blood glucose concentration at physiological conditions.

\begin{figure} [hbt!]
\centerline{\includegraphics[width=30pc]{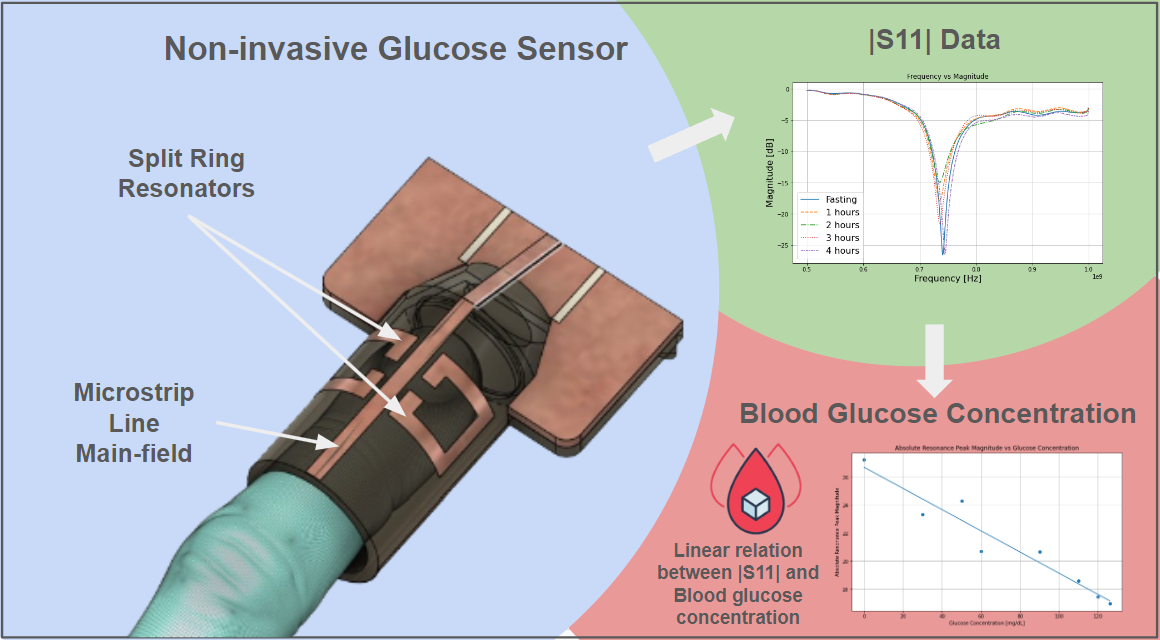}}
\end{figure}

\end{abstract}
\clearpage

\section{Introduction}
Diabetes is a chronic incurable metabolic disease and one of the fastest growing healthcare concerns globally, projected to afflict more than 780 million people globally by 2045\,\cite{hossain2024diabetes}. Early detection and proper management of diabetes rely on regular monitoring of blood glucose levels. However, most commercial glucometers require the invasive collection of blood samples using a small needle for glucose measurements, which causes patients discomfort and damages their nerve system in a long run\,\cite{bruen2017glucose}. In fact, several studies report that the fear of needles can lead to infrequent blood sugar monitoring and result in poorly managed blood glucose in patients with diabetes, highlighting the clinical need for non-invasive glucose measurement devices \cite{al2017fear, cemeroglu2015fear}. This need has spurred recent research into non-invasive glucose measurement devices. Proposed methods include trans-dermal \cite{lipani2018non, sieg2004noninvasive, kost2000transdermal}, optical \cite{liakat2014noninvasive, siu2014plasmonic}, electrochemical \cite{cardosi2012amperometric} and microwave \cite{9262060, kandwal2020highly, oloyo2018highly, 8711633} based devices. 

Most notably, planar microwave based devices have shown promising results owing to the relative low-cost and simplicity of fabrication \cite{9262060}. These sensors are designed to detect the the change in blood and tissue permittivity at different glucose concentrations, by detecting changes in the interactions between the generated fields and the human tissue under measurement using microwave structures such as fringing-field microstrip lines and planar antennas as microwave resonators \cite{oloyo2018highly}. However, these small changes can be extremely challenging to detect in-vivo. High quality factor (Q-factor) sensors feature sharp resonance peaks, small changes in permittivity are more easily detected in the form of frequency shifts or changes in peak magnitudes. Techniques to improve sensitivity include the use of split-ring resonators (SRR) \cite{9262060, oloyo2018highly, 7032431, kandwal2020highly}. For the work above, fringing fields are used, which compromises the sensitivity. There are designs where the main-field of microstrip lines (MLIN) is used for sensing \cite{8476183, huang2018microstrip}. It is shown that higher sensitivity is achieved by using the main fields compared to the fringing fields\,\cite{8476183, huang2018microstrip}. 

Split ring resonators (SRR) are traditionally used to enhance the resonance response in microwave circuits. Typical SRRs found in literature consists of a square or circular ring with a split at one point \cite{abdolrazzaghi2022techniques}. When these structures are placed adjacent to the MLIN, oscillating fields from the MLIN induce oscillating current in the rings. However, the split in the ring interrupts smooth current flow, resulting in electromagnetic resonance in the ring, producing a highly localized and strong resonant field through the substrate, hence increasing the Q-factor of the sensor \cite{salim2016complementary}. However, there is limited research in combining the approaches of main-field MLIN with SRRs to increase  sensitivity. 

In this work, a high-Q, non-invasive blood glucose measurement sensor is proposed, combining the effects of both SRR and main-field MLIN. The proposed sensor design was optimized with simulation, fabricated, and tested both in-vitro (with standard solution phantoms) and in-vivo (with a human subject). In Section II, the sensor design, specifications, simulation and sensing mechanism is discussed in detail. Section III details of the experimental characterization using a phantom and a human study, and the results are presented in Section IV. Section V concludes the paper and outlines the future work.

\section{Sensor Design}
The proposed sensor has a microstrip line (MLIN) flanked by a pair of split ring resonators (SRR) a distance away from the feed, as shown in Figure 1. The sensor is fed by a 50-Ohm coplanar waveguide and is impedance matched.

A shielded housing was designed to contain the entire sensor assembly, reducing the effects of noise from ambient electromagnetic interference. The key dimensions of the sensor are as follows:
\textit{A} = \textit{E} = \textit{F} = 2 mm , \textit{B} = 45 mm, \textit{D} = \textit{C} = 10 mm

\begin{figure} [hbt!]
\centerline{\includegraphics[width=30pc]{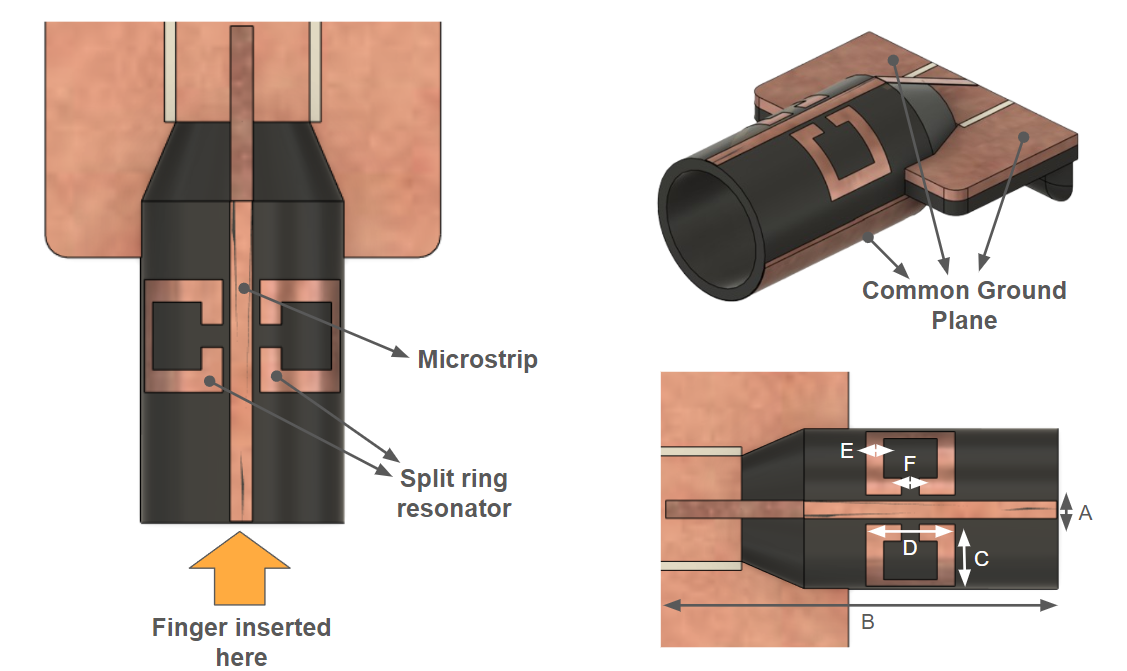}} 
\caption{Proposed sensor design.}
\end{figure}

\subsection{Microstrip Line}
For blood glucose measurements to be taken, the user inserts their fingers into the cylindrical slot as shown in Fig. 1. A typical MLIN consists of a planar signal-carrying line and a ground plane, separated by a substrate material. In this proposed design, the MLIN and the ground plane are positioned at the top and bottom of the cylindrical slot, such that the user's finger serves as the substrate of the line. This method enhances the interactions between the tissue of the fingers and the electric field of the MLIN, as the fields are the so-called main field and are highly confined in the substrate, thus enhancing sensitivity to the changes in glucose concentration in the fingers. 

\subsection{Split Ring Resonators}
The position of the split ring resonators relative to the MLIN must be optimized to maximize the Q-factor of the sensor. There are two possible positional parameters that can be optimized, the distance between the SRR and the feed of the MLIN (X) and the separation of the two rings (Y), as shown in Figure 2(a). The latter is heavily constrained by the dimensions of the cylindrical slot, and is thus fixed as a constant at 16\,mm. To optimize the position along the MLIN, the planar MLIN with SRRs were simulated using Sonnet Lite. Given that the key objective was to optimize the position of planar structures, 2.5D Method-of-Moments (MoM) solver was used as opposed to 3D full-wave solvers to minimize computational resources. A parameter sweep was performed between $X=6mm$ and $X=28mm$, at a step size of 1\,mm. For each value of X, the S11 parameters were calculated and recorded, and the selected S11 plots are given in Figure 2(b). From these results, it shows that $X=16mm$ is the optimal position that produced the greatest Q factor. The sensor with X = 16\,mm and Y = 0.5\,mm was fabricated.  

\begin{figure} [hbt!]
\centerline{\includegraphics[width=30pc]{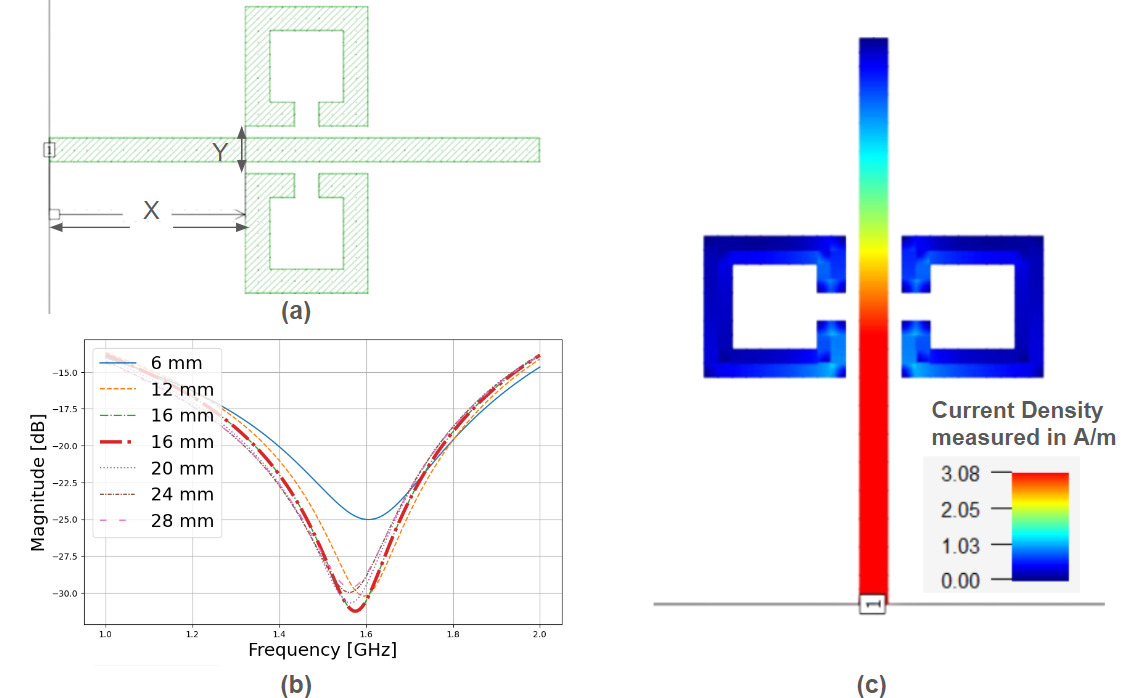}} 
\caption{Simulation optimization (a) Parameter sweep of X used for optimization of SRR position (b) S11 plots for different values of X (c) Surface current density plot of the optimal SRR position}
\end{figure}

\subsection{Sensing Mechanism}
The proposed sensor utilises the resonance of a MLIN-based sensor for sensing, as the reflection at the resonance (called the peak |S11|) increases proportionally to the permittivity of the substrate material. Furthermore, it is known that permittivity decreases as glucose concentration in a given solvent increases \cite{yoon2011dielectric, oloyo2018highly}. The main field sensing method proposed and the above relationship allows for changes in the dielectric properties of the tissues due to the change in blood glucose concentration to be detected as changes in the peak |S11|. This relationship is illustrated as follows. |S11| at the input of the sensor can be expressed as

\begin{equation}
    |S_{11}| = |\Gamma| = \left| \frac{Z_\text{sensor} - Z^\text{CPW}_0}{Z_\text{sensor} + Z^\text{CPW}_0} \right|
\label{eq:S11}
\end{equation}
where $Z_\text{sensor}$ is the input impedance of the sensor and $Z^\text{CPW}_0$ is the characteristic impedance of the CPW, $Z^\text{CPW}_0$\,=\,50\,$\Omega$.  Due to the unique geometry of the proposed strip line, the characteristic impedance can be approximated from the planar microstrip line equation as follows: 

\begin{equation}
    Z^\text{MLIN}_{0} = \frac{a}{\sqrt{\epsilon_{eff}}} \ln\left( \frac{bh}{w} + \frac{w}{ch} \right)
\end{equation}
where $w$ is the width of the microstrip, $h$ is the substrate thickness, the effective dielectric constant, a weighted average of the dielectric constant in air (which is 1) and the dielectric constant in the substrate is given by:

\begin{equation}
    \epsilon_{\text{eff}} = \frac{\epsilon_r + 1}{2} + \frac{\epsilon_r - 1}{2}\left( \frac{1}{\sqrt{1 + 12\left(\frac{h}{w}\right)}} \right)
\end{equation}
For the sensor, it is an open-circuit MLIN with a cylindrical substrate where $Z_\text{in}$ can be expressed as follows:
\begin{equation}
    Z_\text{sensor} = -jZ^\text{MLIN}_0 \beta^\text{MLIN} \ell
\label{eq:Zin}
\end{equation}
where $Z^\text{MLIN}_0$ and $\text{}\beta^\text{MLIN}$ are the characteristic impedance and phase propagation constant of a planar microstrip transmission line, respectively, and $\ell$ is the length of the line. The phase propagation constant is a function of  $ \epsilon_{eff}$  and wave number $k_0$ given by:
\begin{equation}
\beta^\text{MLIN} = k_0 \sqrt{\epsilon_{\text{eff}}}
\end{equation}

The proposed microstrip line geometry was simulated in CST to obtain the Z parameters from a frequency sweep. The mean squared error curve-fitting algorithm was used to estimate coefficients a, b, and c for this design, \textit{a} = -1.590, \textit{b} = 5.808, \textit{c} = -1.815

\vspace{2mm}
Combining all the above relations, the magnitude of S11 can be expressed as a function of the dielectric constant, given as:
\begin{equation}
|S_{11}(\epsilon_r)| = \left| \frac{-\frac{a}{\sqrt{\frac{\epsilon_r + 1}{2} + \frac{\epsilon_r - 1}{2} \left( \frac{1}{\sqrt{1 + 12\left(\frac{h}{w}\right)}} \right)}} \ln\left( \frac{bh}{w} + \frac{w}{ch} \right)j\beta^\text{MLIN} - Z^\text{CPW}_0}{-\frac{a}{\sqrt{\frac{\epsilon_r + 1}{2} + \frac{\epsilon_r - 1}{2} \left( \frac{1}{\sqrt{1 + 12\left(\frac{h}{w}\right)}} \right)}} \ln\left( \frac{bh}{w} + \frac{w}{ch} \right)j\beta^\text{MLIN} + Z^\text{CPW}_0} \right|
\end{equation}

Numerical analysis of the function shows that an increase in the effective dielectric constant results in an increase in the magnitude of the S11 peak, as long as $\mathbf{Z_{sensor} < Z^\text{CPW}_0}$, which is true for the proposed sensor design. Taking into account the above relationships, the magnitude of the S11 peak is inversely related to the blood glucose level.  

\section{Experimental Methods and Materials}
A prototype of the design described in Section II was fabricated and first characterized through the use of standard test solutions with various concentrations of dissolved D-glucose at 2 operating frequencies - 750 MHz and 1.5 GHz. The same prototype was then validated through a human study. 

\subsection{Experimental Measurements using Standard Solution Phantoms}
Standard solutions (0.9\% NaCl) with 8 concentrations of dissolved D-glucose (0, 30, 50, 60, 90, 110, 120, 126 mg/dL) was prepared. In place of a real finger, a synthetic finger phantom with dielectric properties similar to those of human tissue was fabricated using polyvinylpyrrolidone solution \cite{ianniello2018synthesized}. The dimensions of the finger phantom is designed for tight fitting within the cylindrical slot of the sensor, minimizing the effects of air spaces and positional variations of the finger phantom within the sensor. PTFE tubes are embedded within the phantom during fabrication, allowing for standard solutions to be injected. Each concentration of the solution was introduced using a 0.4 mm hypodermic needle until the tubes are completely filled, and the tubes were flushed with deionized water between experiments. The phantom is then inserted into the sensor connected to the experimental set-up as shown in Figure 3. The S11 data were acquired once the reading on the vector network analyzer had stabilized using Instrument View, and processed by Python scripts to extract the maximum peak magnitude and S11 plots.  

\begin{figure} [hbt!]
\centerline{\includegraphics[width= 30pc]{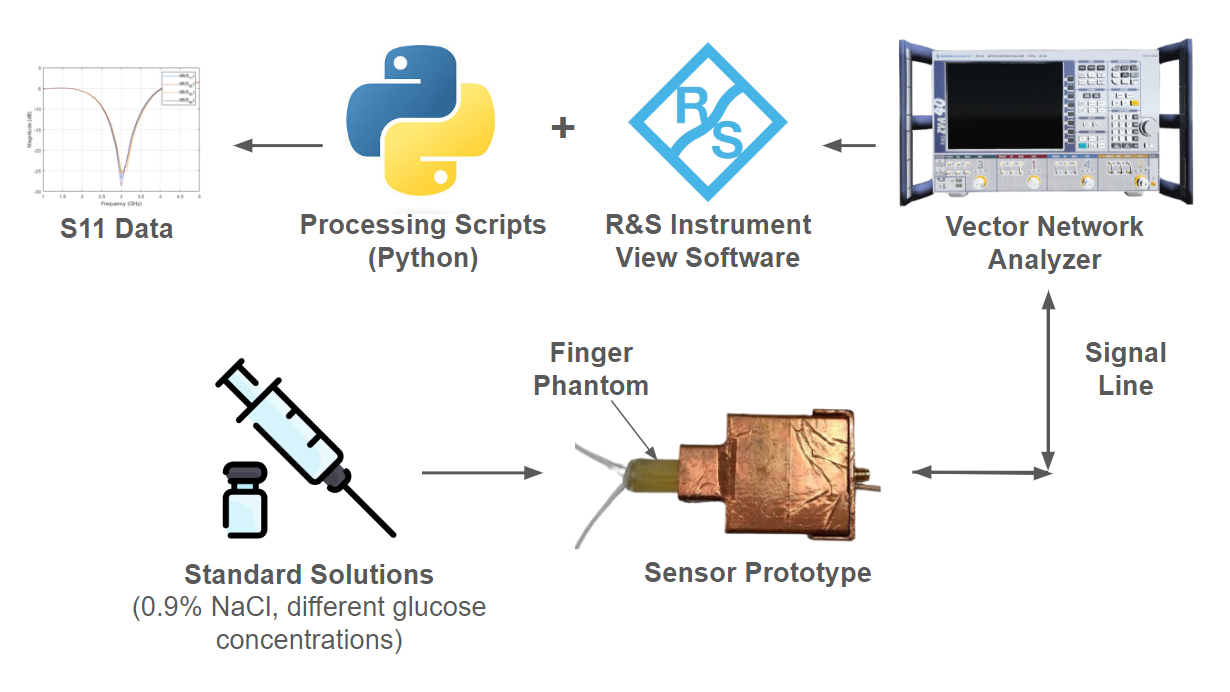}}
\caption{Block diagram of the experimental set up}
\end{figure}

\subsection{In-vivo Measurements in Human Subject}
Following from characterization experiments with standard solution phantoms, the proposed sensor was further tested in-vivo with a human subject at the Biomedical Electromagnetics Lab, Singapore University of Technology and Design. This experiment serves as a first-level validation of the sensor performance in a more realistic operating environment. One healthy male volunteer participated in the study, and informed consent was obtained prior to the experiments. The study was approved by the Singapore University of Technology and Design Institutional Review Board, and the experiment was performed with compliance to the relevant guidelines and regulations. 

The general procedure of this experiment involves measurements taken using the proposed sensor, followed quickly by measurements taken by a glucometer (ACCU-CHEK Performa) on the same finger (left index finger), which served as the ground truth measurement. These measurements were obtained and recorded at 5 instances - fasting, and 1 to 4 hours after a meal. Given that the sensor performance better at 750 MHz operating frequency, the in-vivo experiment was conducted at this frequency. To simulate the operating conditions as far as possible, the amount and composition of the subject's meal was not controlled.  

\section{Results and Discussion}
\subsection{Experimental Measurement Results}
The results from the experimental measurements using standard solutions are summarized in Figure 4. There is an inverse linear relationship between the concentration of glucose and the S11 peak magnitude, which agrees with theoretical predictions. This trend was observed at both the 1.5 GHz and the 750 MHz operating frequency, and both results demonstrate a strong linear coefficient ($R^2 \approx 0.9$). This implies that the proposed sensor is able to distinguish between small changes in glucose concentration through changes in the S11 peak magnitude due to its effect of substrate permittivity. 

\begin{figure} [hbt!]
\centerline{\includegraphics[width=30pc]{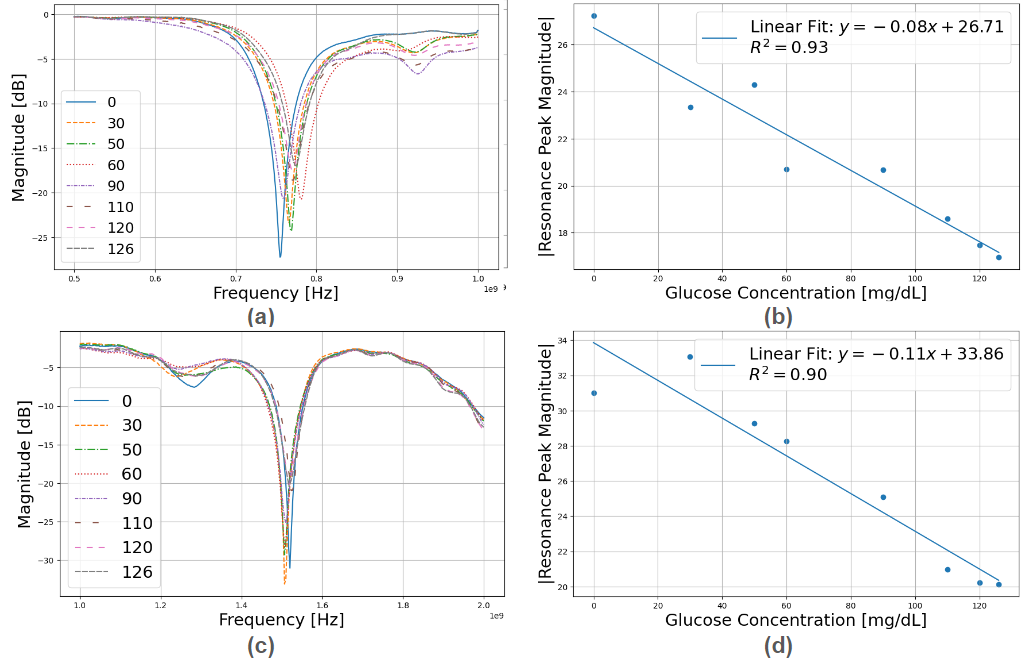}}
\caption{Results from phantom characterization (a) S11 plots at 750 MHz, (b) peak magnitude vs glucose concentration at 750 MHz, (c) S11 plots at 1.5 GHz, (d) peak magnitude vs glucose concentration at 1.5 GHz.}
\end{figure}

\subsection{Human Study Results}
The results of the human study is summarized in Figure 5. The peak magnitudes of the S11 plots in Figure 5(a) correspond to the blood glucose concentrations of the subject measured using the glucometer at different time intervals after meals in Figure 5 (b). The observed trend is similar to the trend demonstrated experimentally in Section IIIA, further validating the working principle and performance of the proposed sensor. 

\begin{figure} [hbt!]
\centerline{\includegraphics[width=30pc]{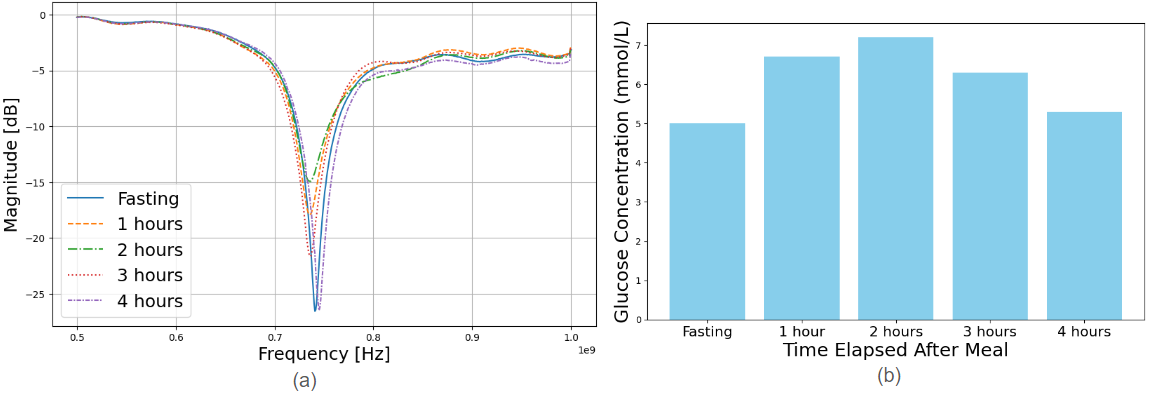}}
\caption{Results from in-vivo human study (a) S11 plots  (b) Blood glucose concentration from glucometer}
\end{figure}

The results from this paper have demonstrated that the proposed sensor is sensitive enough to take advantage of the nearly inverse linear relationship between the S11 peak magnitude and the blood glucose concentrations in the microwave frequency range for non-invasive glucose measurement. Given that sensitivity to small changes in glucose concentration remains a major limitation to developing microwave-based noninvasive glucose sensors, the authors believe that the findings of this work are significant. 

\section{Conclusion} 
In this work, a high-Q non-invasive glucose sensor utilising microstrip line main field and split ring resonators was developed. Simulation was used to optimize the design, and the optimized sensor geometry was fabricated and validated. Based on the change in S11 peak magnitude due to change in substrate permittivity, the proposed sensor demonstrated the sensitivity to distinguish between changes in glucose concentration within the physiological range, through the standard solution experiments and the in-vivo human study. At both 1.5 GHz and 750 MHz operating frequency, the same highly linear inverse relationship between S11 peak magnitude and glucose concentration was observed, which could potentially be used for calibration purposes. 
Moving forward, the proposed sensor should be validated on a greater sample size of human subjects. The extensive data set obtained could be used in the training of a machine learning classifier model. Such data-driven models will be able to utilize more parameters and discover hidden patterns in the S11 data, revealing the potential for a more robust sensor that could be a suitable candidate for monitoring glucose in patients with diabetes.

\section*{Acknowledgment}
The authors would like to thank John Koh from SUTD STEM Labs for providing chemicals, lab equipment and support for the experiments, without which the development, testing and validation of the proposed sensor would not have been possible.

\bibliographystyle{IEEEtran}
\bibliography{References}\

\end{document}